\documentclass[aps,prl,twocolumn,showpacs,floatfix]{revtex4}
\usepackage{graphicx}
\usepackage{hyperref}
\usepackage{amsmath}
\usepackage{psfrag}

\def\>{\rangle}
\def\<{\langle}

\begin{document}

\title{Resource-efficient linear optical  quantum computation}
\author{Daniel E. Browne\footnote{Present address: Clarendon Laboratory, University of Oxford, Parks Road, Oxford OX1 3PU, UK} and Terry Rudolph}
\address{QOLS, Blackett Laboratory, Imperial College London,
Prince Consort Road, London SW7 2BW, UK}

\begin{abstract}
We introduce a scheme for linear optics quantum computation, that
makes no use of teleported gates, and requires stable interferometry
over only the coherence length of the photons. We achieve a much
greater degree of efficiency and a simpler implementation than
previous proposals. We follow the ``cluster state'' measurement
based quantum computational approach, and show how cluster states
may be efficiently generated from pairs of maximally polarization
entangled photons using linear optical elements. We demonstrate the
universality  and usefulness of generic parity measurements, as well
as introducing the use of redundant encoding of qubits to enable
utilization of destructive measurements - both features of use in a
more general context.
\end{abstract}

\pacs{03.67.Lx,03.67.Mn,42.50.Dv}

\maketitle

\section{Introduction}
Our understanding of the sufficient requirements for quantum
computation has been greatly enhanced by Knill, Laflamme and
Milburn's \cite{klm} (KLM) discovery that measurement induced
nonlinearity suffices for efficient quantum computation.
Specifically they showed that linear optical elements (beam splitters,
phase shifters etc) combined with single photon sources and single
photon detectors can, in principle, be used for efficient quantum
computation. In practice, even given these resources, significant
obstacles stand in the way of making the KLM scheme a feasible technology
for quantum computation. These include: (i) The sheer number of
optical elements required, (ii) a need for extremely good, and very
large, quantum memory (iii) a requirement of keeping what is
essentially a giant interferometer phase stable to within a photon
wavelength.

In this article we present a theoretical proposal for quantum
computation with photons and linear optics which, in addition to a
considerable number of other advantages, either overcomes or greatly
alleviates all these key issues. We then demonstrate a core module
of this proposal experimentally. Our proposal moves completely away
from the use of teleportation to boost non-deterministic gate
probabilities. Rather, we introduce two ``fusion'' mechanisms, which
allow for the construction of entangled photonic states, known as
cluster states. These states, introduced by Briegel and Raussendorf
\cite{briegelcluster}, allow for universal quantum computation by
performing single qubit measurements \cite{clusterqc}. Since arbitrary single-qubit
measurements are easy to perform on photonic qubits, it follows that
our construction enables efficient quantum computation.

One key advantage of using cluster states is that the quantum gates
are implemented with unit probability, rather than the
``asymptotically unit'' probability of the original KLM scheme.
Other proposals to avoid this feature of the KLM scheme were
presented by Yoran and Reznik \cite{reznik} and Nielsen
\cite{nielsencluster}; the latter also made use of cluster states.
However both of these proposals make use of the same fundamental
teleportation primitives introduced by KLM, and thus suffer similar
problematic features. In contrast, our proposal overcomes the
issues of non-deterministic gate operations by introducing the use
of what we call ``redundant encoding'' of qubits.

The primary resource we will make use of is two photon Bell
states. These can be obtained in a purely via linear optics and photo-detection
with probability 3/16 from four single photons \cite{inprep}. In fact, since any
non-trivial non-deterministic gate will create some entanglement,
which can then be purified if necessary
\cite{jianwei_pur}, a wide variety of options for creating this
initial resource exist. Alternatively, it is also quite feasible
that non-linear optical processes be used create the initial
entanglement \cite{sliwabanaszek}. Given the Bell states, we
proceed to build up the cluster states using only
non-deterministic parity-check measurements~--~which involve
combining the photons on a polarizing beam-splitter (PBS) followed immediately by measurement
on the output modes.

In addition to overall smaller resource requirements in terms of
number of single photons, linear optical elements and measurements
required (we estimate factors of several orders of magnitude over
Nielsen and many orders of magnitude over KLM, since the entangled resource states they require are generated via several or many low probability non-deterministic operations), our proposal has several other advantages.
First, if we
are prepared to accept a small (constant factor) overhead in
resources, a simple extension of our basic proposal also has the
significant advantage that photon-number-discriminating detectors
are \emph{not} required for its implementation.

Moreover, there is no requirement for elaborate interferometers containing
multiple beam splitters in series, which greatly reduces the
complexity of mode-matching issues in an experimental
implementation. More dramatically, it also removes the  requirement
of maintaining the phase stability of an extremely large and complex
interferometer. The non-deterministic gates introduced by KLM, which are also the basis of \cite{reznik} and \cite{nielsencluster}, rely on Mach-Zehnder-type interference, which is sensitive to pathlength phase instabilities on the order of the photon's wavelength,  i.e.  around a micrometer for infra-red light.
In contrast, the interference we make use of is of the simple
Hong-Ou-Mandel ``coincidence'' form, and thus only requires
stability over the coherence length of the photons, a much larger distance. Recent down-conversion experiments \cite{panfour} have obtained  coherence lengths on the order of 10{$^{-4}$} m and in quantum dot experiments \cite{yoshi} coherence lengths several orders of magnitude greater than this have been reported. Thus the basic component of our scheme is at least three orders of magnitude less sensitive to phase instability than previous proposals. 

\begin{figure}
\psfrag{aa}{a)}
\psfrag{bb}{b)}
\psfrag{cc}{c)}
\begin{center}\includegraphics[width=6cm]{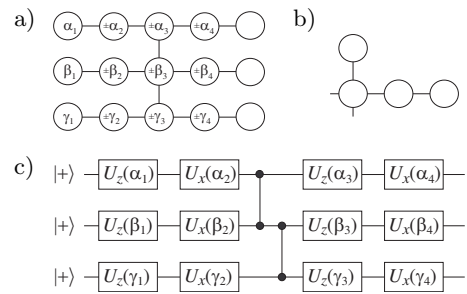}\end{center}
\caption{\label{nielsencluster} The  measurement pattern (a) simulates
the quantum network (c). Each
circle represents a qubit in the cluster and each line represents
a ``bond'' - i.e. a CPHASE having been applied between the two
connected qubits. The observable $\cos(\theta\/)\sigma_x +
\sin(\theta\/)\sigma_y$ is measured on each qubit, with the angle
$\theta$ given each time by the symbols inside the circle. The
sign of the measurement angle in all but the first column depend
upon the outcome of measurements to the left of the qubit. Larger
circuits can be simulated by larger cluster states with extensions
of this pattern. Such layouts can be generated by tiling repeated
3-bond units of the ``L-shape'' shown (b). }
\end{figure}

Let us review the salient features of cluster state computation
\cite{clusterqc} phrased in terms of photonic qubits. Special
entangled states, known as cluster states \cite{briegelcluster} are
generated by applying a controlled-phase shift (CZ gate) between
nearest neighbors in a square lattice of qubits initially in the
superposition of horizontal and vertical polarization state
$|H\rangle+|V\rangle$ (all states are unnormalized). The
next-neighbor entanglement can be paraphrased  as a ``bond'' between
the qubits and thus the layout of the bonds which define the state
can be represented graphically, as, for example, in
Fig.~\ref{nielsencluster}. Once the cluster state is generated, a
quantum logic network is implemented by measuring the qubits
individually in a particular pattern of measurement eigenbases and
in a particular order. Given a cluster state of sufficient size, any
quantum circuit can be implemented, and the states are thus an
entanglement resource for universal quantum computation.
Applying  Pauli unitary operators to any subset of qubits in the cluster
state creates a different but computationally equivalent cluster state, as when the
measurements for the computation are implemented the extra Pauli's
are taken into account in the measurement pattern
\cite{clusterqc}. We thus consider all cluster states which
are related by operations in the Pauli group  as equivalent; they
are represented by the same layout of nearest-neighbor bonds.

The square lattice cluster state of Raussendorf and Briegel's
original scheme is extremely powerful, allowing the  simulation of
unitaries directly without decomposing them into a network of some
set of gates. However, if one wishes to minimize the number of
inter-qubit bonds in the cluster, a different approach is more
appropriate. To simulate a quantum network made up of arbitrary
rotations and controlled-phase gates, the  cluster state layout in
Fig.~\ref{nielsencluster}  (suggested by Nielsen in
\cite{nielsencluster}, although his scheme cannot actually realise its most compact form) is sufficient, and requires far fewer
inter-qubit bonds. In this paper we will concentrate on generating
cluster states with this more compact layout.

\begin{figure}
\psfrag{aa}{a) Type-I}
\psfrag{bb}{b) Type-II}
\begin{center}\includegraphics[width=6cm]{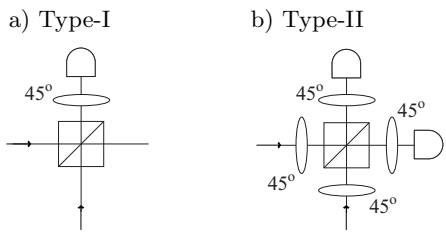}\end{center}
\caption{\label{parity} The two
``qubit fusion'' non-deterministic gates: (a) Type-I consists of a
mixing the two spatial modes on a polarizing beam splitter (PBS),
which reflects vertically polarized light only, and measuring one
of the output modes with a polarization discriminating photon
counter after a 45$^\circ$ polarization rotation; b) Type-II is
obtained from Type-I by adding both 45$^\circ$ rotations to
each input mode and  measuring the output modes in the rotated
basis. The 4 polarization rotators and PBS could be replaced by a
PBS rotated at 45$^\circ$. }
\end{figure}

First we first describe a ``qubit fusion'' operation which is very
important for our proposal. This parity-check
\cite{jianwei_pur,franson} operation is implemented by mixing the
two modes on a polarizing beam splitter (PBS), rotating one of the
output modes by 45$^\circ$  before measuring it with a
polarization discriminating photon counter (see
Fig.~\ref{parity}(a)). Since we introduce a second fusion
operation later, we refer to this as Type-I fusion. (Type-I fusion has some parallels with
the valence-bond solid interpretation of cluster states \cite{frank}.)

\begin{figure}
\psfrag{aa}{a)}
\psfrag{bb}{b)}
\begin{center}\includegraphics[scale=0.3]{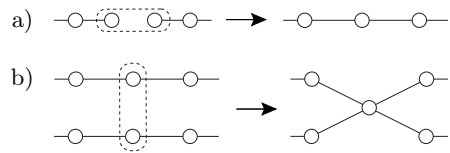}\end{center}
\caption{\label{fig:fuse} Two examples of how clusters states can  be
joined together by ``fusing''  qubits from each. a) Fusing the end-qubits
of two linear clusters of length $n$ and $m$  creates a cluster of length
$(n+m-1)$. b) The mid qubits of two linear clusters are fused to create a
two-dimensional cluster with a cross-like layout. In this way non-trivial
cluster layouts can be created.}
\end{figure}

The effect of this operation depends upon the outcome of the measurement. Let us assume
that the input state had at most one photon in each spatial mode. In this case, when one
and only one photon is detected (which occurs with probability 50\% for the cluster state
inputs we need to consider), the state is transformed by the Kraus operators
$(|H\rangle\langle HH|- |V\rangle\langle VV|)/\sqrt{2}$ or $(|H\rangle\langle HH|+
|V\rangle\langle VV|)/\sqrt{2}$ depending on whether a horizontally or vertically
polarized photon is detected. The aptness of the name ``fusion'' becomes apparent when
one considers the effect this has when applied to two qubits in separate cluster states.
Since the CZ operation is diagonal in the computational basis $\{|H\rangle,|V\rangle\}$,
the ``fused'' qubit inherits all the cluster state bonds of the two qubits which were
fused (see Fig.~\ref{fig:fuse}, cf. \cite{frank}). If the Type-I fusion is applied to the end-qubits of
linear (i.e. one-dimensional) clusters of lengths $n$ and $m$, successful outcomes
generate a linear cluster of length $(n+m-1)$ (Fig.~\ref{fig:fuse}(a)). Note that the two
successful outcomes generate equivalent cluster states.

\begin{figure}
\psfrag{aa}{a)}
\psfrag{bb}{b)}
\psfrag{cc}{c)}
\begin{center}\includegraphics[width=6cm]{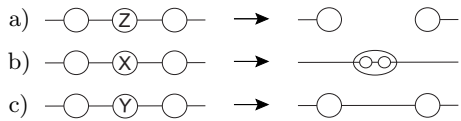}
\end{center}
\caption{\label{clusident}Certain measurements
on a cluster qubit will leave the remaining qubits in a new
cluster state with a different layout: a) A $\sigma_z$ eigenbasis
measurement removes the qubit from the cluster and breaks all
bonds between that qubit and the rest of the cluster. b) A
$\sigma_x$ measurement on a linear cluster removes the measured
qubit and causes the neighboring qubits to be joined such that
they now represent a single logical qubit with logical basis
$|00\rangle,|11\rangle$. c) A $\sigma_y$ measurement removes the qubit from the linear cluster but links the neighboring qubits. These gain an extra $\pi/2$ rotation around the $z$-axis which is accounted for when they are measured.}
\end{figure}

The Type-I fusion  operation is considered to have failed when
either zero or two photons of either polarization are detected.
The failure outcomes are described by Kraus operators
$|0\rangle\langle VH|$ and $(|2_V\>-|2_H\rangle)\langle
HV|/\sqrt{2}$, and  have the effect of measuring both input qubits
in the $\sigma_z$ eigenbasis (the computational basis). Measuring a cluster state qubit in the
computational basis leaves the remaining qubits in a cluster state
of the same layout as before the measurement, but now with all the
bonds connected to the measured qubit severed (see
Fig.~\ref{clusident}(a)).

Starting from a supply of polarization Bell states (which are equivalent to a 2-qubit
cluster state $|HH\rangle+|VH\rangle+|HV\rangle-|VV\rangle$), the Type-I fusion
operation allows us to efficiently generate arbitrarily long linear cluster states. In
the simplest case, a single successful Type-I fusion combines two Bell pairs into a
3-qubit cluster state, (which is also a GHZ state). Since, on average, one must attempt
this whole procedure twice before the desired three-qubit cluster is generated, the
expected number of Bell states used to generate the 3-qubit cluster state is 4. We shall
use the ``expected number of Bell states consumed'' as a measure of the resources
required to generate cluster states of a given size.

A simple strategy to generate a long linear cluster  is
to first generate an intermediate supply of 3-qubit cluster
states, and then attempt to fuse these one by one to a larger
linear cluster. Each time, with probability $1/2$, the cluster
grows in length by 2 qubits, or, equally likely, loses a qubit. A
failed attempt creates a Bell pair from the 3-qubit cluster, which
can be reused in the generation of further 3-qubit clusters. Thus,
on average, the cluster grows by $1/2$ a qubit in length for each
attempt, and the resources needed scale as $(2\times4-1)=7$
Bell pairs per qubit in the linear cluster. (The subtracted amount
represents the 2-qubit clusters which can be reused). A more
efficient method is to first generate 5-qubit clusters by
combining 3-qubit clusters. Since failures leave 2-qubit clusters
which can be reused, the mean resources required to create a
5-qubit cluster are 14 Bell pairs. To utilize these in creating
arbitrary length clusters one may do the following: One attempts
to add the 5-qubit cluster, if the fusion fails one then tries to
attach the 4-qubit cluster which is generated, if it fails again a
3-qubit cluster is created which can be reused to generate further
5-qubit clusters. Taking this recycling into account, the mean
resources needed with this method are 6.5 Bell pairs per qubit
added to the linear cluster. We do not know the optimal procedure
for generating the linear clusters by Type-I fusion.

\begin{figure}
\includegraphics[width=6cm]{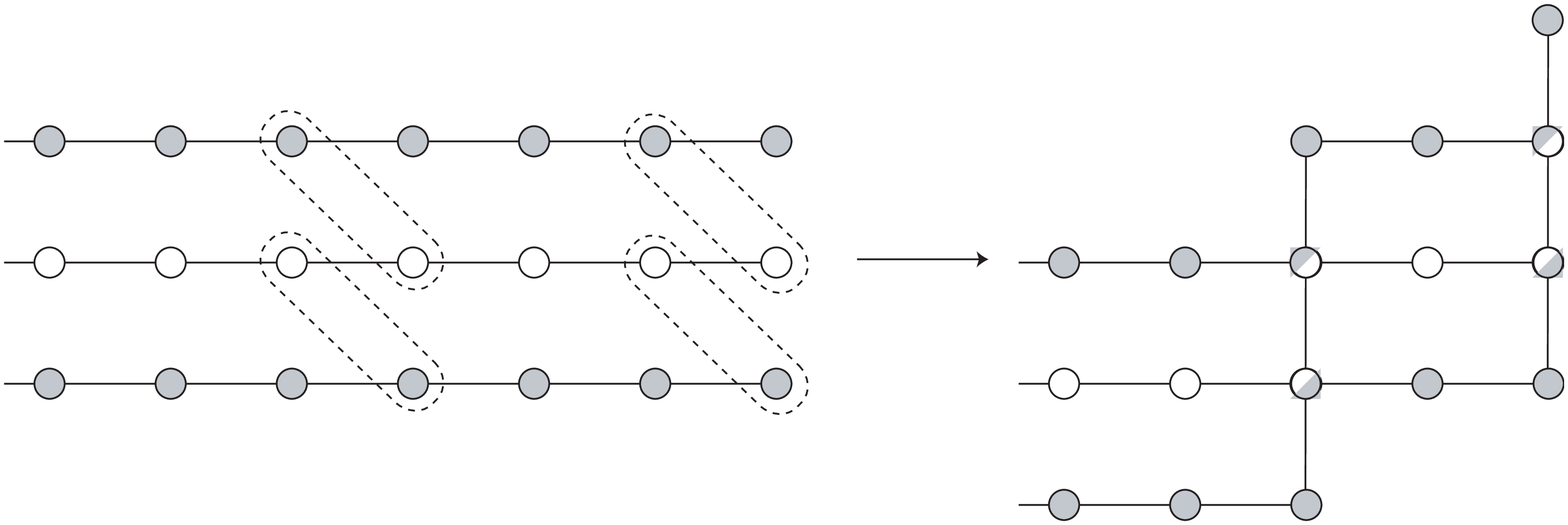}
\caption{\label{meldlayout} If qubits from a linear cluster are fused
according to the above pattern, a cluster state with the desired layout is generated.}
\end{figure}

One-dimensional  clusters are not, however, sufficient for
universal quantum computation, as their geometry doesn't permit
the implementation of 2-qubit gates.  We thus need to create
two-dimensional clusters, which can also be done by fusion, as
depicted in Fig.~\ref{fig:fuse}(b). More precisely, we envisage
fusing together qubits in  linear clusters, as is illustrated in
Fig.~\ref{meldlayout}, which shows how the  layout from
Fig.~\ref{nielsencluster} can be achieved.

Type-I fusion operation  is not appropriate for carrying out these fusions, since its
failure outcome is a measurement in the computational ($\sigma_z$) basis, which would
split the linear clusters in two (Fig.~\ref{clusident}(a)).
Another approach to fusion is clearly necessary. In this alternate approach we introduce
the use of \emph{redundant encoding}. A single qubit in the cluster may be represented by
multiple photons, such that a generic cluster state $|\phi_0\>|0\>+|\phi_1\>|1\>$ could
be encoded $|\phi_0\>|H\>^{\otimes n}+|\phi_1\>|V\>^{\otimes n}$, where we have singled
out from the rest of the cluster the particular qubit which is redundantly encoded with
$n$ photons. Note that a $\sigma_x$ measurement (projection onto $|H\>\pm|V\>$) on one of
the redundant photons does not destroy the cluster state, it removes one photon from the
redundant encoding and perhaps adds an inconsequential phase.

A $\sigma_x$ measurement also has an interesting effect when
performed on a qubit in a linear cluster; it does not split the
cluster, rather it combines the adjacent qubits into a single
redundantly encoded (by two photons) qubit, retaining the bonds
attached to each, as shown in Fig.~\ref{clusident}(b).

To utilize these features of $\sigma_x$ measurements, we make use of the gate depicted in
Fig.~\ref{parity}(b). When it succeeds, with probability 1/2, (as heralded by the
detection of a  photon in each output mode) this gate is a destructive projective
measurement onto maximally entangled states, i.e. the Kraus operators are
$(\<HH|+\<VV|)/\sqrt{2}$, $(\<HV|+\<VH|)/\sqrt{2}$. The failure outcome (signaled by
detecting no photons in one of the modes) effectively performs a projective measurement
of $\sigma_x$ on each of the photons. Note that the Type-II fusion does not require the
discrimination between different photon numbers. 

We see therefore, that if this gate is applied to a single photon
of each of a pair of logical qubits in the redundant $n$-photon
encoding, it will lead to the desired fusion. If it fails then one
photon is removed from each qubit's redundant encoding, and the
gate could be reattempted, as long as sufficient photons remained
in each qubits redundant encoding.

\begin{figure}
\psfrag{aa}{a)}
\psfrag{bb}{b)}
\psfrag{cc}{c1)}
\psfrag{dd}{c2)}
\psfrag{success}{\emph{success}}
\psfrag{failure}{\emph{failure}}
\begin{center}\includegraphics[width=7cm]{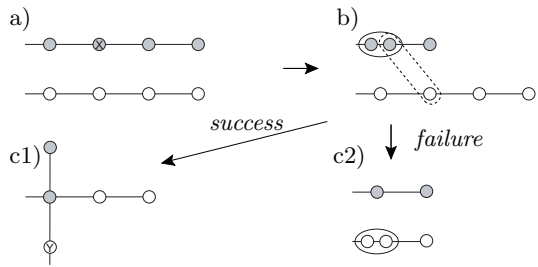}\end{center}
\caption{\label{averagefusion} Here we illustrate the  construction of the ``L-shape'': a) A $\sigma_x$-measurement causes the neighboring qubits to be
joined into a single logical qubit in the redundant encoding. b)
Type II-fusion is now attempted between this logical qubit and a
qubit in the lower cluster. The fusion succeeds with probability
1/2. c1) If the fusion succeeds, a single further $\sigma_y$ measurement creates the desired L-shape (see Fig. 4c). c2) If it fails, a redundantly encoded qubit is created on the lower cluster. The qubits are now in a pattern similar to step b, so with the addition of two further qubits another Type-II fusion can be attempted. These steps are repeated until a successful fusion is accomplished. On average, creating the L-shape uses up 8 bonds from the linear clusters involved.}
\end{figure}

It turns out that this gate works even when one of the logical qubits is represented by
two photons, and the second by just a single photon, since these operators take the state
$(|\phi_0\rangle|HH\rangle+ |\phi_1\rangle|VV\rangle)\otimes (|\xi_0\rangle|H\rangle+
|\xi_1\rangle|V\rangle)$ to $|\phi_0\rangle|H\rangle|\xi_0\rangle+
|\phi_1\rangle|V\rangle|\xi_1\rangle$ and $|\phi_0\rangle|H\rangle|\xi_1\rangle+
|\phi_1\rangle|V\rangle|\xi_0\rangle$ respectively, which are both the desired ``fused''
cluster states. We call this a Type-II fusion. The effect  of the failure outcome of the
Type-II fusion is to perform a $\sigma_x$ measurement on each photon. This has the
consequence of converting the redundantly encoded 2-photon logical qubit into a 1-photon
logical qubit on the one cluster, while creating a new redundantly encoded 2-photon qubit
on the lower linear cluster (see Fig.~\ref{averagefusion}). Thus the fusion attempt can
be immediately re-attempted. The mean number of times that the fusion must be attempted
is simply $\sum_{n=1}^\infty (1/2)^nn=2$.

Cluster states  with the layout illustrated in
Fig.~\ref{nielsencluster} can be  generated by combining the two
processes outlined above, i.e. first  generating of linear
clusters by Type-I fusion, and then fusing their qubits by Type-II
fusion to form the desired 2-dimensional cluster.

We can quantify  the resources required to build the cluster by
recognizing that the layout of Fig.~\ref{nielsencluster} can be
broken down into the L-shaped units   illustrated in
Fig.~\ref{nielsencluster}(b). Thus, the resources to
construct such a L-shape gives an appropriate way of quantifying
the resources required per two-qubit gate in the logical network.
The L-shape can be constructed from two linear clusters via a
single (successful)  Type-II fusion. A method of generating the L-shape is illustrated in Fig.~\ref{averagefusion}. On average, two Type-II fusion attempts are required and  8 qubits bonds from the linear
clusters involved are used up. Note that unlike in \cite{nielsencluster}, there is no back-propagation of errors here  into the already generated cluster, meaning that the cluster qubits can be measured as soon as the next adjacent L-shape has been completed.
Since  constructing
the linear clusters requires on average no more than 6.5 Bell pairs for each
qubit in the cluster, construction of the L-shape requires on average no more than 52 Bell pairs. This is a great improvement compared with other linear
optics based quantum computation schemes of which the authors are
aware \cite{klm,reznik,nielsencluster}.

For instance, the most efficient scheme so far is Nielsen's
approach in
\cite{nielsencluster}. Remember that each attempt of the
implementation of a KLM CZ$_{n^2/(n+1)^2}$ gate requires a 4n-photon entangled state for its implementation.
Nielsen calculates  that 24 successful  CZ$_{4/9}$ gates are
required per implemented two-qubit logical gate. Considering the
number of times that a gate with success probability $(4/9)$ must
be repeated, we see that in Nielsen's scheme $24 \times \frac{9}{4}=54$ 8-photon entangled states are consumed per two-qubit gate.
These  8-photon entangled states must be generated via a very
complicated non-deterministic procedure involving multiple
beam-splitters and non-deterministic gates (see
\cite{klmlongpaper}). In our simpler scheme, the resources per logical 2-qubit
gate in our network are the same as the resources used to add a ``L-shape'' to
the cluster, on average 52 Bell pairs.


We have made minimal use of the redundant encoding introduced for
Type-II fusion. In fact, by using a redundant encoding for all
qubits in a cluster it is possible to use only the parity gate of
Fig.~\ref{parity}(b) for \emph{all} gate operations. This has the
considerable advantage that the gate can be implemented
\emph{without} photon number discriminating detectors, and naturally detects photon absorption errors.
Since, in this case, two photons would be measured in each fusion, Bell states would not be a sufficient initial resource, one would have to use  three-photon cluster states instead, which increases the resource requirements by a constant factor. 
 The nature
of such a redundant encoding also allows for a single qubit to simultaneously be involved
in bonding operations with multiple (possibly widely separated) other qubits. Incidentally, CZ gates 
(as opposed to fusion operations) between redundantly encoded qubits can be directly implemented
via the gate of
Fig.~\ref{parity}(b), with an extra 45$^\circ$ rotation on one input mode. 

We have introduced a scheme for linear optical quantum computation which has significantly lower
resource requirements than previous proposals, and would be far less demanding in terms of phase stability.
Although we have phrased our results in terms of photon polarization,
parity measurements are a natural 2-qubit measurement in bosonic
systems. 
In fact, there
has been much interest in the general question of when measurements
can replace (all or part) of the processes of the standard circuit
model. Our results can be interpreted as contributing to this effort
by providing the first proof that parity measurements (even
non-deterministic ones), combined with single qubit
transformations/measurements, are universal for quantum computing.


We thank Viv Kendon, Paul Kwiat, Chao-Yang Lu, Jeremy O'Brien, Jian-Wei Pan, Kevin Resch, Petra Scudo and ``ceptimus''
of the puzzles forum at randi.org for helpful comments. TR would
like to thank Michael Nielsen for early discussions about his
ideas, the Dept. of Physics at University of Queensland for their
hospitality, and in particular Andrew White, whose constant
discourse on `mode-matching' formed a primary source of irritation
for this work. This research was supported by the EPSRC and
Hewlett-Packard Ltd.



\begin{thebibliography}{99}
\bibitem{klm} E. Knill, R. Laflamme, and G. Milburn, Nature  (London) {\bf 409}, 46 (2001).
\bibitem{briegelcluster} H.J. Briegel and R. Raussendorf, Phys. Rev. Lett. {\bf 86}, 910-913 (2001).
\bibitem{clusterqc}  R. Raussendorf and H. J. Briegel, Phys. Rev. Lett. {\bf 86}, 5188-5191 (2001); R. Raussendorf, D.E.~Browne and H.J.~Briegel, Phys. Rev. A {\bf 68}, 022312 (2003).
\bibitem{reznik} N. Yoran and B. Reznik, Phys. Rev. Lett. {\bf 91}, 037903 (2003).
\bibitem{nielsencluster} M.A. Nielsen,  Phys. Rev. Lett. {\bf 93} 040503 (2004).
\bibitem{inprep} Simple Procrustean linear optical schemes for the generation of multi-party entangled states (including Bell states, GHZ states, etc.) from single photons can be constructed by mixing pairs of perpendicularly polarised single photons at a beam splitter and applying fusion gates to filter out the desired states.


\bibitem{jianwei_pur} J.-W. Pan \textit{et al.}, Nature (London) \textbf{423}, 417 (2003).
\bibitem{sliwabanaszek}  C. \'Sliwa and K. Banaszek, Phys. Rev. A {\bf 67}, 030101(R) (2003).


\bibitem{panfour} J.-W. Pan \textit{et al.}, Phys. Rev. Lett. \textbf{86}, 4435 (2001).
\bibitem{yoshi} C. Santori \textit{et al.}, New J. Phys. \textbf{6}, 89 (2004).
\bibitem{franson}     T. B. Pittman, B. C. Jacobs, and J. D. Franson,  Phys. Rev. A {\bf 64}, 062311 (2001).
\bibitem{frank} F.~Verstraete and J.I.~Cirac, Phys. Rev. A 70, 060302(R) (2004).

\bibitem{klmlongpaper} E. Knill, R. Laflamme, and G. Milburn, e-print quant-ph/0006088.

\end{thebibliography}
\end{document}